\documentclass[preprint,12pt]{elsarticle}

\usepackage{amssymb}
\usepackage{amsmath}
\usepackage{subcaption}
\usepackage{mhchem}
\usepackage{graphicx}
\usepackage{float}
\usepackage{natbib}
\usepackage[colorlinks=true,linkcolor=black, citecolor=blue, urlcolor=blue]{hyperref}
\usepackage[T1]{fontenc}
\usepackage{mathptmx}
\usepackage{mathtools}
\usepackage{hyperref}
\usepackage{booktabs,chemformula}
\usepackage{cuted}
\usepackage{lipsum}
\usepackage{mwe}
\usepackage{amsthm}
\usepackage{amsfonts}
\usepackage{footnote}
\usepackage[labelformat=simple]{subcaption}

\journal{Nuclear Physics B}

\begin{document}

\begin{frontmatter}



\title{Effects of neutron radiation on the optical and structural properties of blue and green emitting plastic scintillators}


\author[a]{V.~Baranov}
\author[a]{Yu.I.~Davydov}
\author[b,c]{R.~Erasmus}
\author[d]{C.O.~Kureba}
\author[e]{N.~Lekalakala}
\author[e]{T.~Masuku}
\author[e]{J.E.~Mdhluli}
\author[e,f]{B.~Mellado}
\author[d]{G.~Mokgatitswane\corref{cor1}}
\ead{gmokgatitswane@gmail.com}
\author[e,c]{E.~Sideras-Haddad}
\author[a]{I.~Vasilyev}
\author[g]{P.N.~Zhmurin}

\cortext[cor1]{Corresponding author}
\address[a]{Joint Institute for Nuclear Research, Dubna, Russia, 141980}
\address[b]{School of Physics, University of the Witwatersrand, Johannesburg, Wits 2050, South Africa}
\address[c]{DST-NRF Centre of Excellence in Strong Materials, University of the Witwatersrand, Johannesburg, Wits 2050, South Africa}
\address[d]{Department of Physics and Astronomy, Botswana International University of Science and Technology, Private Bag 16, Palapye, Botswana}
\address[e]{School of Physics and Institute for Collider Particle Physics, University of the Witwatersrand, Johannesburg, Wits 2050, South Africa}
\address[f]{iThemba LABS, National Research Foundation, PO Box 722, Somerset West 7129, South Africa}
\address[g]{Institute for Scintillation Materials, Kharkov, Ukraine}

\begin{abstract}
We report on the optical and structural properties of plastic scintillators irradiated with neutron beams produced by the IBR-2 reactor of the Frank Laboratory of Neutron Physics in JINR, Dubna. Blue UPS-923A and green plastic scintillators were irradiated with neutron fluence ranging from 10$^{13}$ to 10$^{17}$ n/cm$^2$. Discolouring in the plastic scintillators was observed after irradiation. The effects of radiation damage on the optical and structural properties of the samples were characterized by conducting light yield, light transmission, light fluorescence and Raman spectroscopy studies. The results showed that neutron radiation induced damage in the material. The disappearance of the Raman peak features in green scintillators at frequencies of 1165.8, 1574.7 and 1651.2 cm$^{-1}$ revealed significant structural alterations due to neutron bombardment. Losses in fluorescence intensity, light yield and light transmission in the plastic scintillators were observed.

\end{abstract}

\begin{keyword}
Neutron radiation \sep Neutron fluence \sep Radiation damage \sep Scintillator \sep Polystyrene 
\end{keyword}

\end{frontmatter}


\section{Introduction}
\label{intro}
Plastic scintillators are employed within high energy particle detectors due to their desirable properties such as high optical transmission and fast rise and decay times~\cite{Knoll:1300754}. The generation of fast signal pulses enables efficient data capturing. They are used to detect the energies and reconstruct the path of the particles through the process of luminescence due to the interaction of ionising radiation. Compared to inorganic crystals, plastic scintillators are organic crystals that are easily manufactured and therefore cost effective when covering large areas such as the ATLAS detector~\cite{chen}.\\ 
\indent In the ATLAS detector of the Large Hadron Collider (LHC), there is a hadronic calorimeter known as the Tile Calorimeter that is responsible for detecting hadrons, taus and jets of quarks and gluons through the use of plastic scintillators. These particles deposit large quantities of energy and create an immensely detrimental radiation environment. The neutrons mostly coming from the shower tails contribute to the counting rates and degradation of plastic scintillators through neutron capture. Monte-Carlo calculations have been performed to estimate doses and particle fluences currently experienced at different regions of the detector, operating at a nominal luminosity of 10$^{34}$ cm$^{-2}$s$^{-1}$. The maximum neutron fluence per year in the Tile Calorimeter barrel was estimated at around 10$^{12}$ n/cm$^{2}$yr~\cite{Angela}. The LHC intends to increase its luminosity by a factor of up to ten times by 2022, and this will drastically impact the radiation environment in the ATLAS detector.\\        
\indent The interaction of ionising radiation with plastic scintillators results in the damage of these plastic scintillators. According to Sonkawade {\it {et al}}.~\cite{sonkawade}, during irradiation the properties of scintillators are altered significantly depending on the structure of the target material, fluence and the nature of radiation. Some of these structural modifications have been ascribed to the scissoring of the polymer chain, intensification of cross-linking, breakage of bonds and formation of new chemical bonds. This damage results in a significant decrease in the light yield of the scintillator and as a result, errors are introduced in the data captured.\\
\indent Studies on proton irradiated plastic scintillators conducted by the Wits High Energy Physics (Wits-HEP) group have been reported in literature~\cite{SAIP, 1742-6596-645-1-012019, NIMB}. This paper extends the study to focus on the effects of non-ionising radiation (i.e. neutrons). Compared to the interaction of ionising radiation, the interaction of non-ionising radiation with matter is more interesting since the particles interact indirectly with the atoms of the material. When materials are bombarded with neutrons, collision cascades are created within the material that results in point defects and dislocations. A Primary Knock-on Atom (PKA) is created when the kinetic energy from the collision is transferred to the displaced lattice atom. The knock-on atoms lose energy with each collision and that energy in turn ionizes the material~\cite{Bisanti}. Neutron irradiation allows for bulk probing on materials since they are highly penetrating particles.

\section{Experimental Details}
\label{methods}
Commercial blue scintillators UPS-923A~\cite{sci} and recently synthesized green scintillators \cite{velmo,yu} were investigated. The samples were prepared at the Institute for Scintillation Materials (ISMA, Kharkov). They were cut and polished to dimensions of 2 x 2 cm with 6 mm thickness. Table~\ref{table:properties} has some important properties of the plastic scintillators under study.

\begin{table}[H]
	\begin{center}
		\caption{Properties of the scintillators under study.}
		\label{table:properties}
		\begin{tabular}{lll}\hline
			Scintillator& Blue UPS-923A & Green\\ \hline
			Manufacturer& Institute for Scintillation& Institute for Scintillation\\
			& Materials& Materials\\
			Base& Polystyrene & Polystyrene\\
			Primary fluor& 2$\%$ PTP& 3HF\\
			Secondary fluor& 0.03$\%$ POPOP\\
			Light Output & 60\\
			\,\,\,\,($\%$ Anthrance)&&\\
			Wavelength of & 425& 530\\
			\,\,\,\,\,\,\,Max. Emission (nm)&&\\
			Rise time (ns)& 0.9& 0.9\\
			Decay time (ns)& 3.3& 7.6\\ \hline 			
		\end{tabular}	
	\end{center}
\end{table}

\begin{figure}[H]
	\begin{center}
	  \begin{subfigure}[normla]{0.49\textwidth}
	  \includegraphics[scale=0.42]{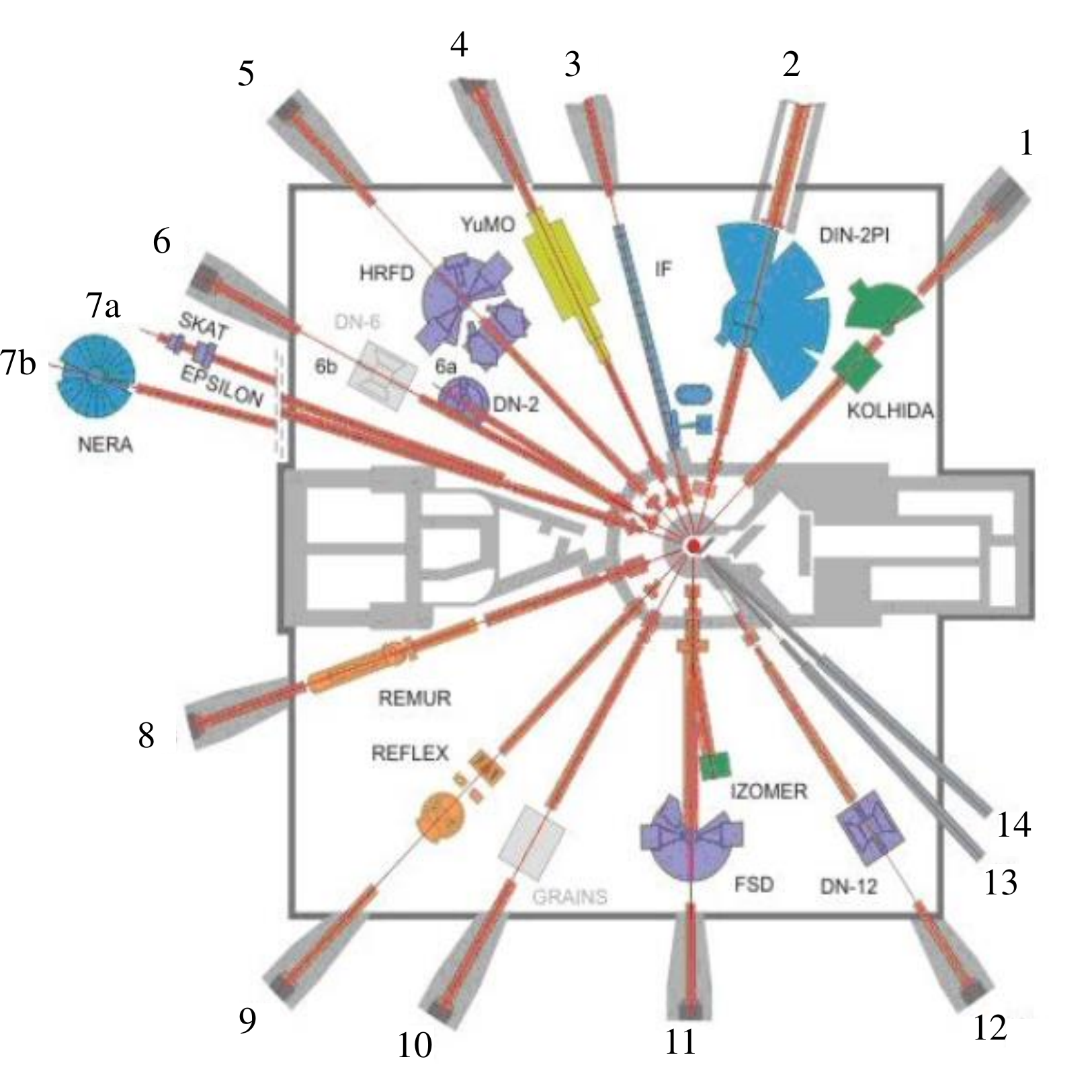}
	  \caption{}
	  \label{subfig:a}
	  \end{subfigure}
	   \begin{subfigure}[normla]{0.49\textwidth}
	  \includegraphics[scale=0.42]{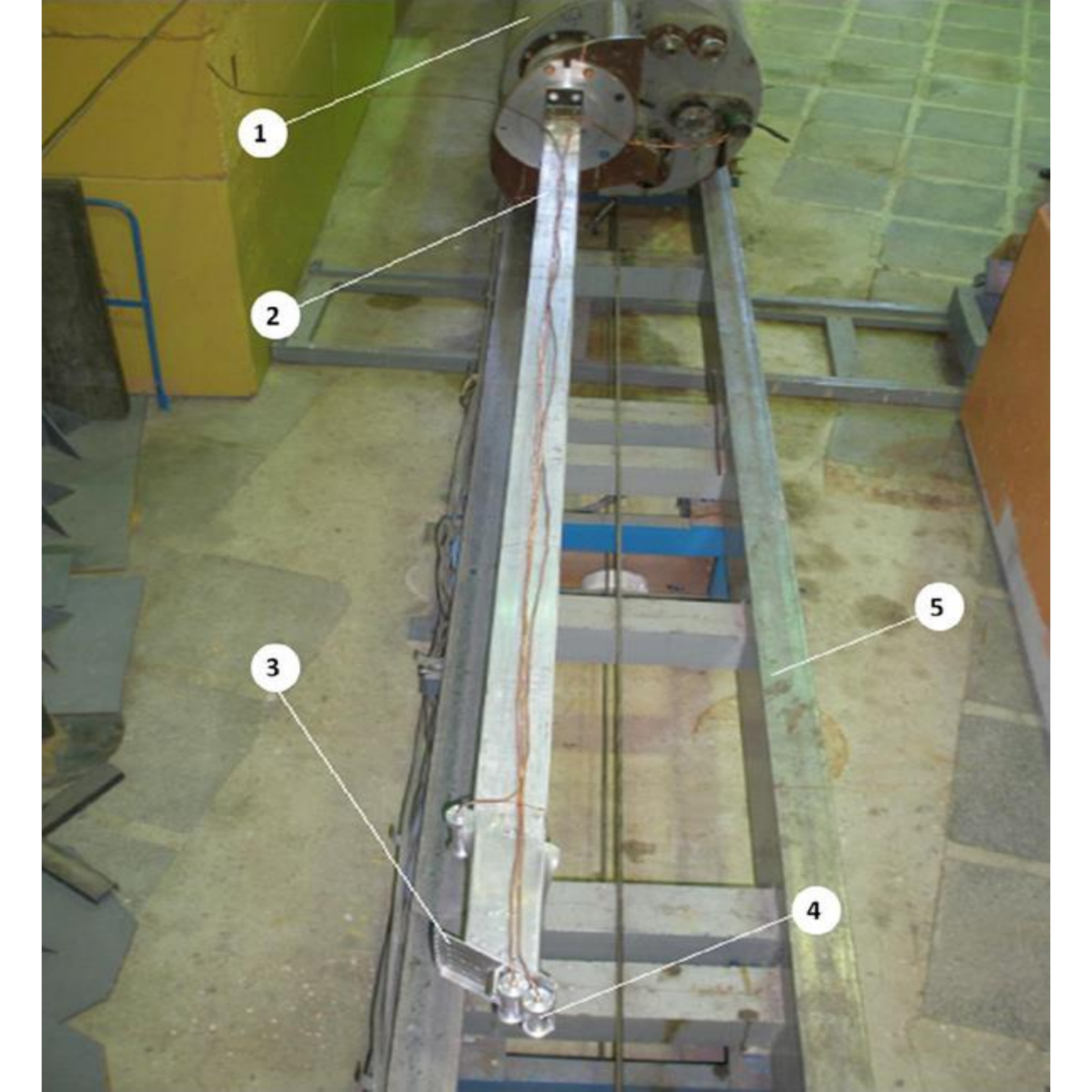}
	  \caption{}
	  \label{subfig:b}	  
	  \end{subfigure}
	  \caption[Layout of IBR-2 spectrometer complex~\subref{subfig:a}, and the irradiation facility at the channel No. 3 of IBR-2 reactor experimental hall, the view from the external biological shield side: (1)-massive part of the irradiation facility, (2)-transport beam, (3)-metallic container for samples fastening, (4)-samples, (5)-rail way]{Layout of IBR-2 spectrometer complex~\subref{subfig:a}, and the irradiation facility at the channel No. 3 of IBR-2 reactor experimental hall, the view from the external biological shield side: (1)-massive part of the irradiation facility, (2)-transport beam, (3)-metallic container for samples fastening, (4)-samples, (5)-rail way~\subref{subfig:b}~\cite{bulav}.}
     \end{center}
\end{figure}

Channel number 3 of the IBR-2 reactor, as schematically shown in Figure~\ref{subfig:b}, located at the Frank Laboratory of Neutron Physics (FLNP) at the Joint Institute for Nuclear Physics (JINR) in Dubna, Russia was used to irradiate the samples~\cite{bulav, shabalin}.

\begin{figure}[H]
	\begin{center}
		\begin{subfigure}[normla]{1\textwidth}
			\includegraphics[scale=0.54]{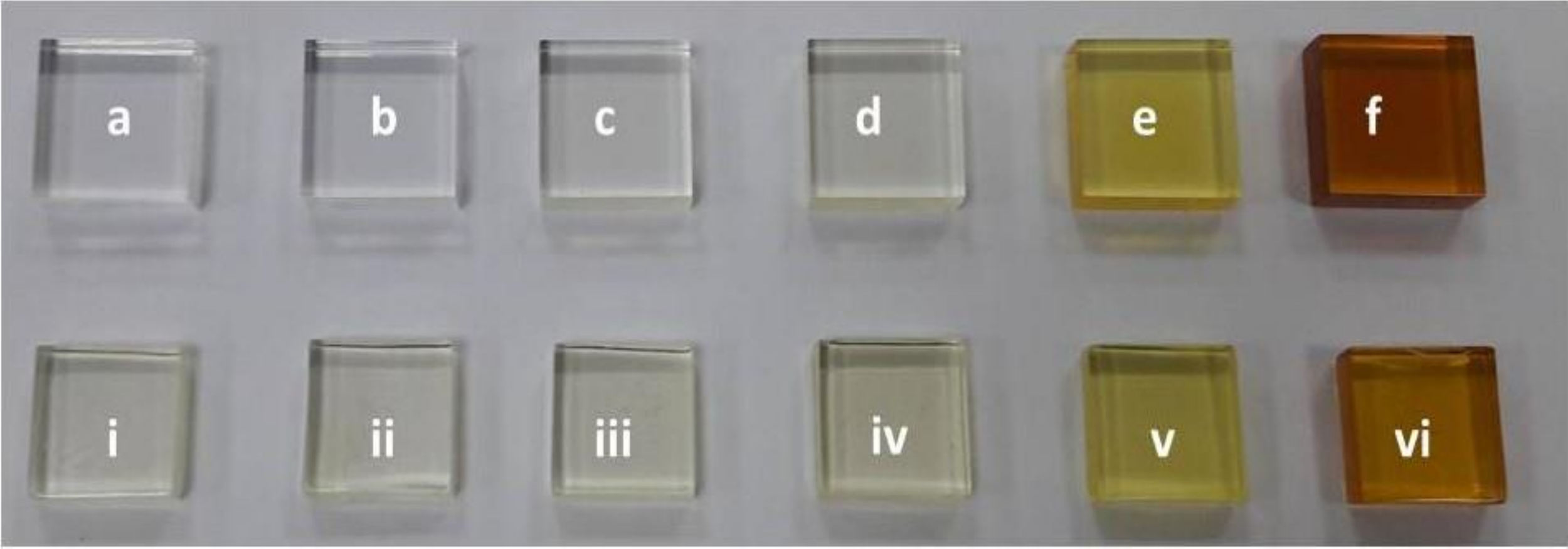}
		\end{subfigure}
		\caption{Neutron irradiated samples, Blue scintillators UPS-923A (a - f) and Green scintillators (i - vi). From the left column to right column: non-irradiated,  10$^{13}$, 10$^{14}$, 10$^{15}$, 10$^{16}$ and 10$^{17}$ n/cm$^2$.}
	\label{fig:samples}	
	\end{center}
\end{figure}

The samples were subjected to a beam of neutrons for 432 hours, this was the duration of the reactor cycle during the October 2017 run (9 $-$ 27 October 2017). The reactor operated at an average power of 1875 kW with the samples placed at various positions away from the reactor core to achieve various neutron fluences. The neutron fluence ranged approximately between 10$^{13}$ $-$ 10$^{17}$ n/cm$^2$. However, during the irradiation only neutrons with energy $E>1$ MeV were monitored. These neutrons account for about a quarter of the total flux. Hereafter we refer to number of fast neutrons with energy $E>1$ MeV, although the actual amount of neutrons are a factor of four higher. As shown in Figure~\ref{fig:samples}, the discolouration of the samples is evident after irradiation. 

\begin{figure}[H]
	\begin{center}
		\begin{subfigure}[normla]{0.68\textwidth}
			\includegraphics[height=8cm, width=10cm]{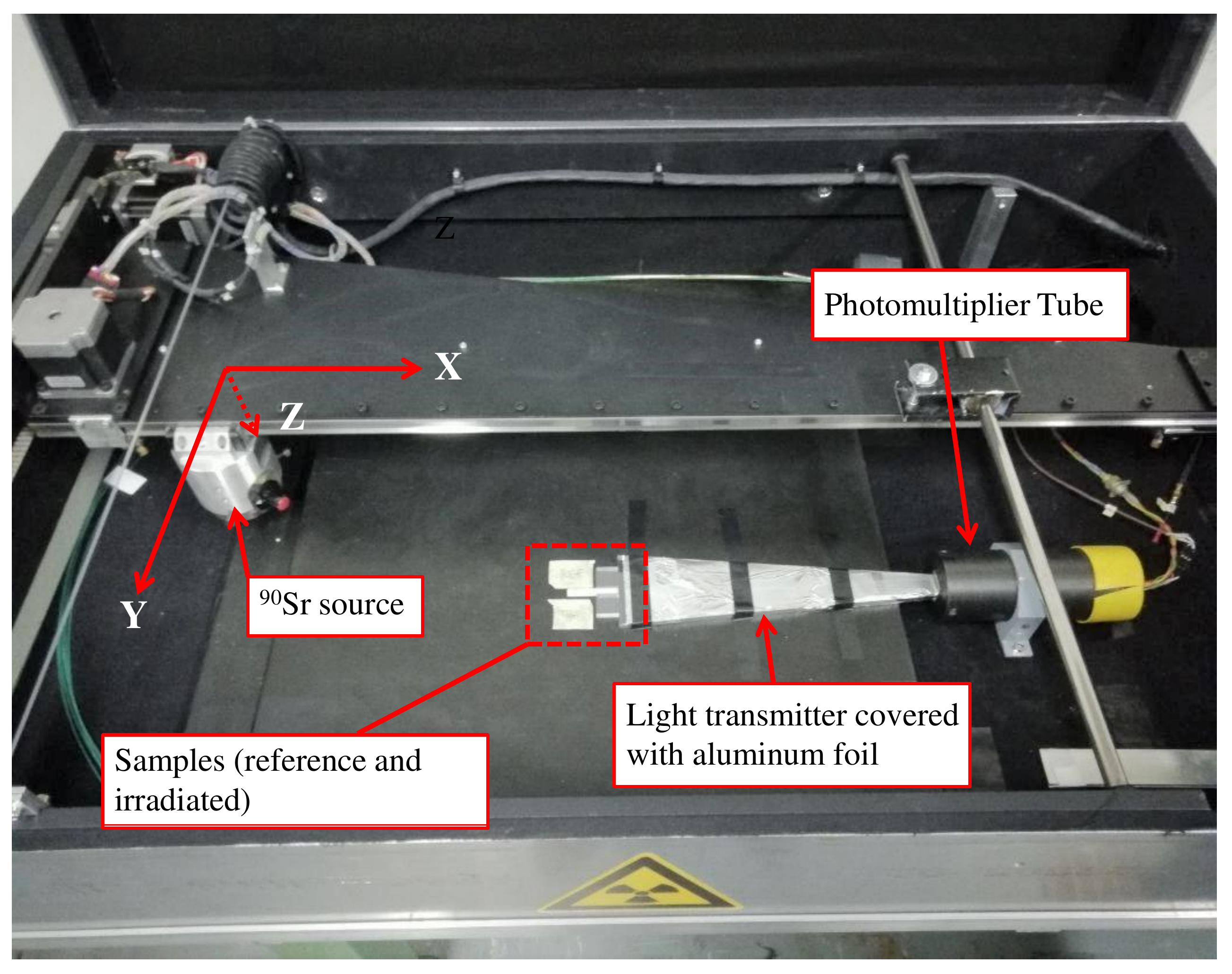}
		\end{subfigure}
		\caption{A photograph of light box set-up used to measure the light yield.}
		\label{fig:box}
	\end{center}
\end{figure}

The studies of effects of radiation damage on the optical and structural properties of the samples were characterized by conducting light yield, light transmission, light fluorescence and Raman spectroscopy measurements. Transmission spectroscopy studies were conducted using the Varian Cary 500 spectrophotometer located at the University of the Witwatersrand. Light transmission was measured relative to transmission in air over a wavelength range of 300-800 nm. The spectrophotometer consists of a lamp source and diffraction grating to produce a differential wavelength spectrum of light. A tungsten lamp was used to produce light in the visible spectrum and a deuterium lamp was used to produce light in the ultra-violet spectrum.\\
\indent Light fluorescence measurements of the neutron irradiated plastic scintillators were conducted at the University of the Witwatersrand using the Horiba LabRAM HR Raman spectrometer. Light emission resulting from the luminescence phenomenon was excited in the plastic scintillators using a laser excitation wavelength ($\lambda$$_{ex}$) of 244 nm, operating at a power of $\sim$20 mW. A laser spot size of 0.7 $\mu$m provided energy for molecular excitations to occur. A grid of 11 x 11 points (121 acquisition spots) was mapped across a surface area of 200 x 200~$\mu$m using a motorised X-Y stage. This allowed for an average representative spectrum to be determined largely free from local variations introduced by surface features such as scratches. \\ 
\indent The light yield measurements were conducted at the European Organisation for Nuclear Research (ATLAS-experiment) using a light tight box set-up shown in Figure~\ref{fig:box}. The plastic scintillators were excited with $\beta$-electrons emitted by a \ce{^{90}Sr} source with average energies of 0.54 MeV and 2.28 MeV. The \ce{^{90}Sr} source scanned over the sample in the X$-$Y direction whilst emitting radiation in the Z direction. The light emitted by plastic scintillators through fluorescence was detected by the photomultiplier tube (PMT). The signal generated by the PMT was further processed through electronics and digitized. To minimize background signals like those coming from the interaction of the $\beta$-electrons with the PMT, a light transmitter was used to transport the light produced by the scintillators to the cathode of the PMT. In addition, a light transmitter was covered with aluminum foil to impede $\beta$-electrons from the source.   

\indent Structural properties of the plastics were characterized using Raman spectroscopy at the University of the Witwatersrand. The Horiba LabRAM HR Raman spectrometer was used to obtain the Raman spectra for the non-irradiated control samples as well as the irradiated samples. A 785 nm diode laser was used to excite the Raman modes and the spectrograph was calibrated via the zeroth order reflection of a white light source from the grating.

\section{Results and Discussion}
\subsection{Raman Spectroscopy Results and Analysis}
Raman spectroscopy measurements were performed with the aim of assessing changes in the structure and morphology of the irradiated samples. Raman spectra were obtained for the irradiated and non-irradiated samples using Horiba LabRAM HR Raman spectrometer, with a laser excitation wavelength of 785 nm. This laser wavelength was chosen as it was found that with a green excitation wavelength the higher fluence samples gave a very prominent background fluorescence that was of sufficient intensity to mask the Raman peaks. With the longer excitation wavelength there were no problems with background fluorescence and hence better quality spectra were obtained. 

\begin{figure}[H]
	\begin{center}
		\begin{subfigure}[normla]{0.49\textwidth}
			\includegraphics[scale=0.26]{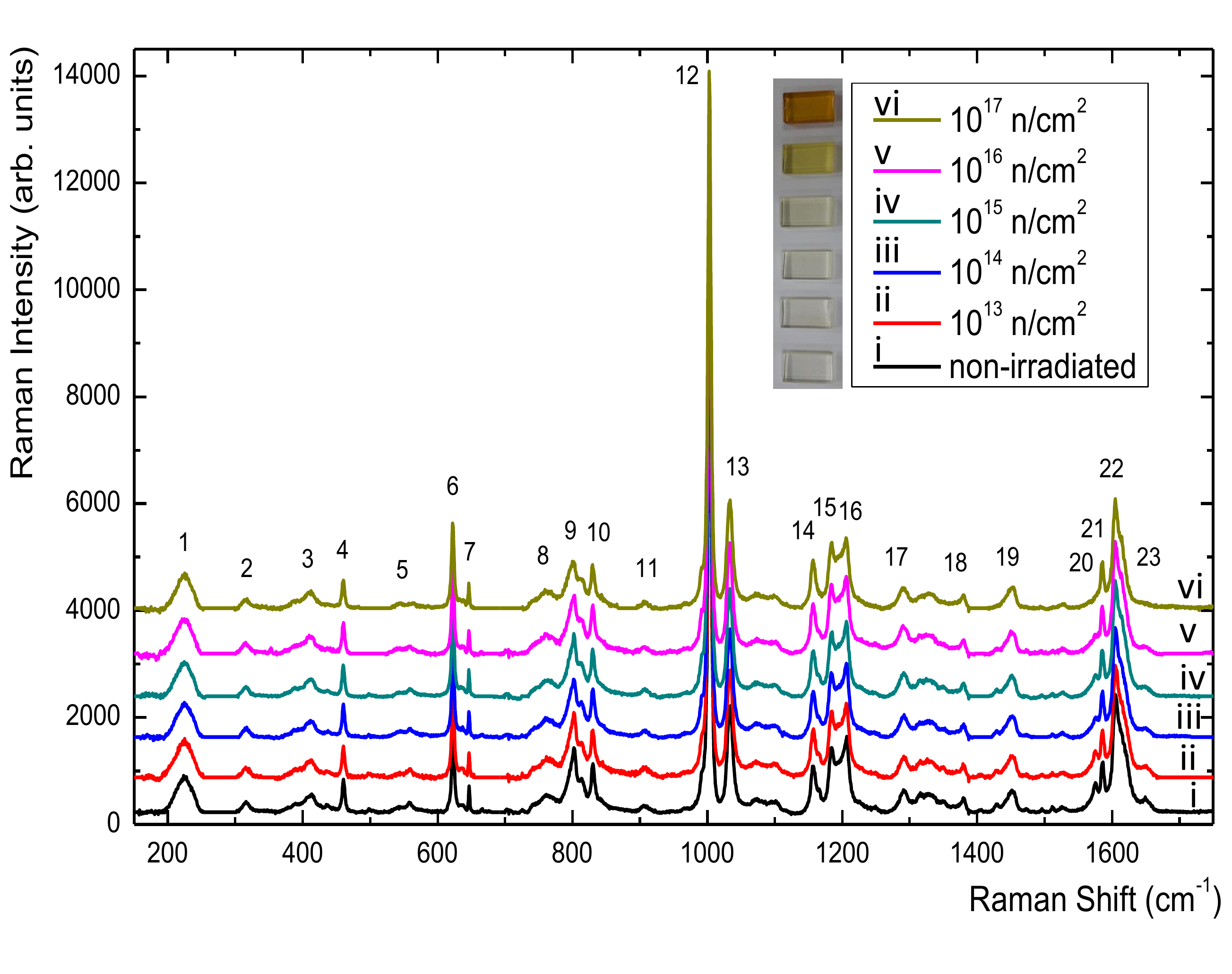}
			\caption{}
			\label{subfig:aramspec}
		\end{subfigure}
		\begin{subfigure}[normla]{0.49\textwidth}
			\includegraphics[scale=0.26]{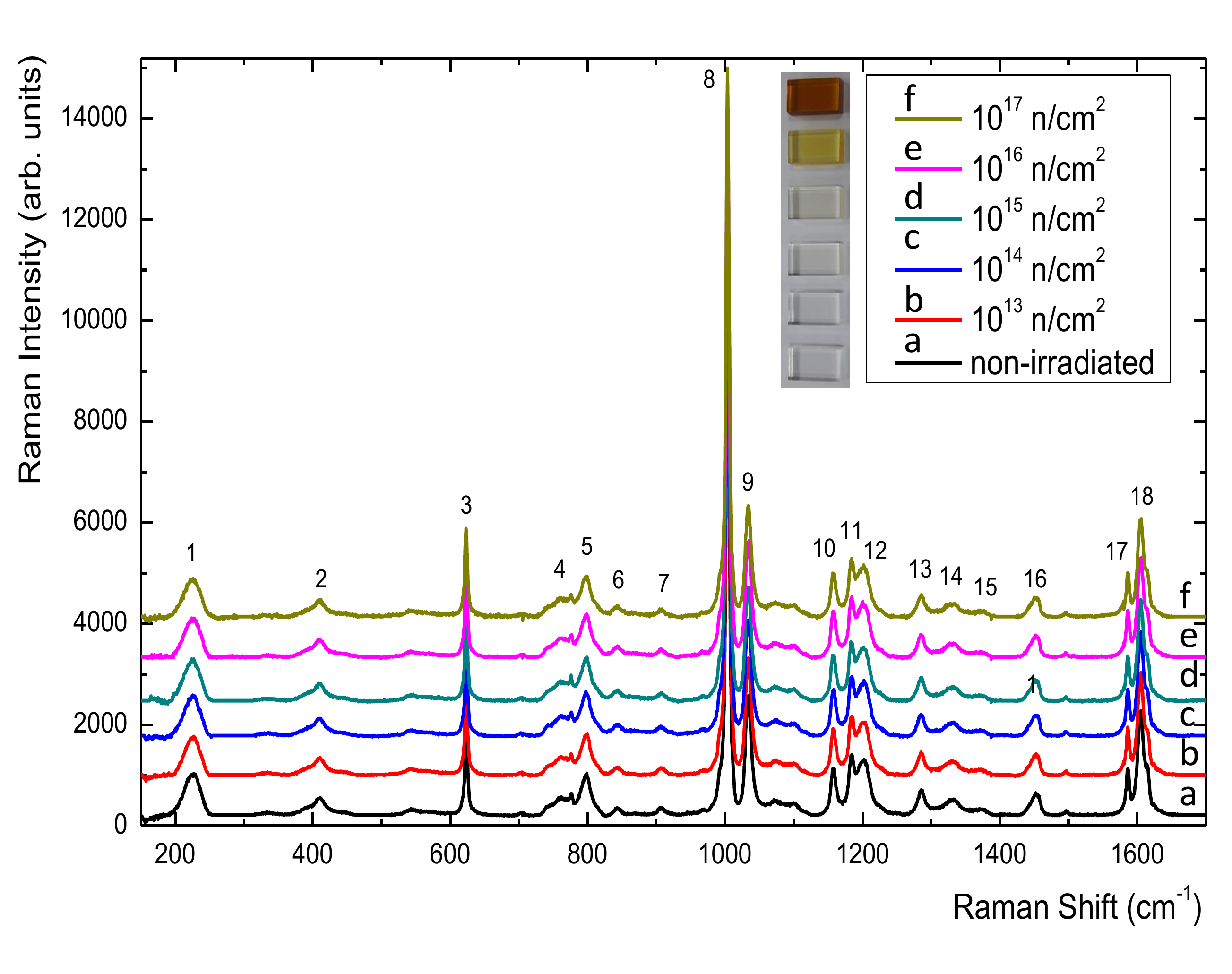}
			\caption{}
			\label{subfig:bramspec}	  
		\end{subfigure}
		\begin{subfigure}[normla]{0.49\textwidth}
				\includegraphics[scale=0.26]{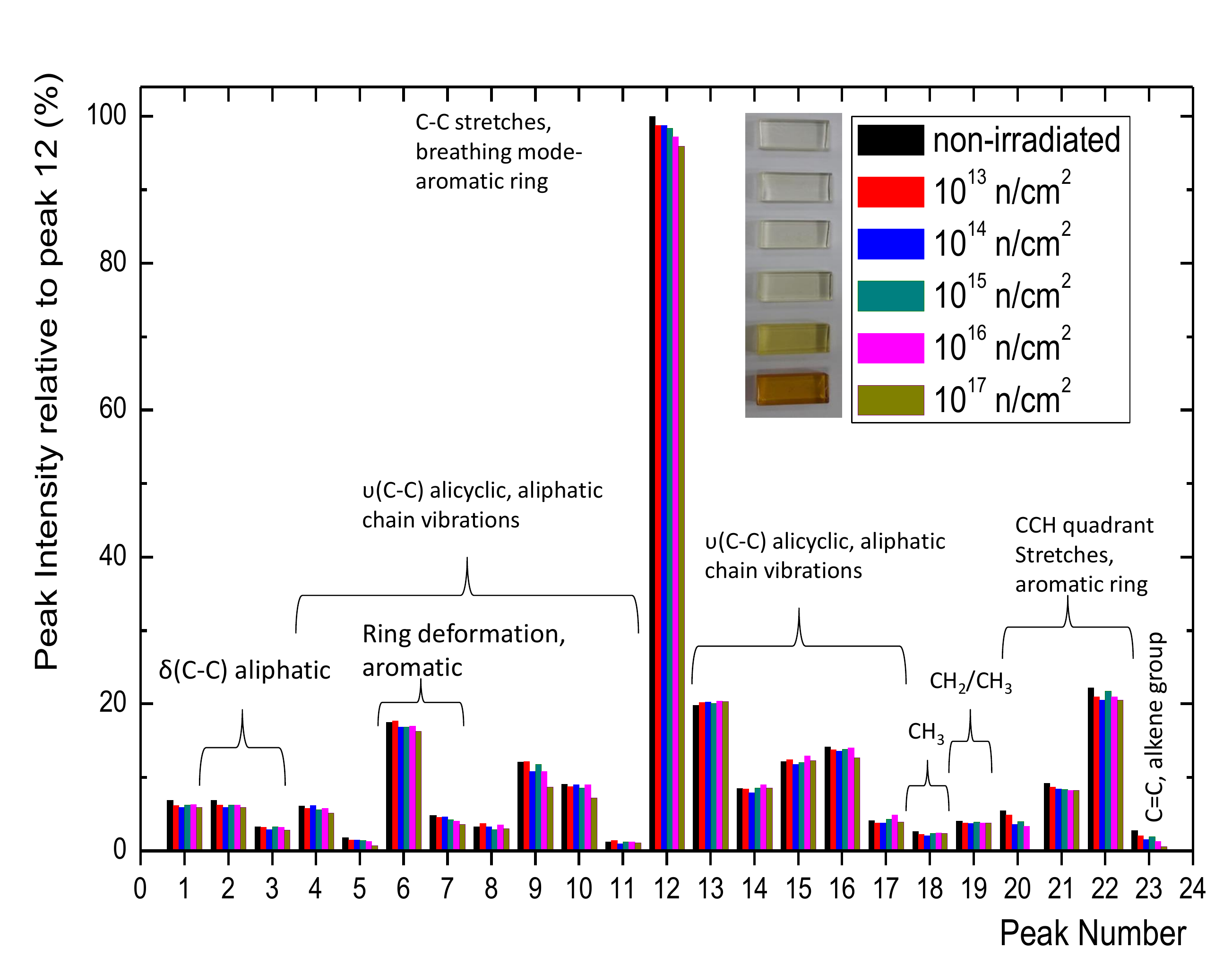}
				\caption{}
				\label{subfig:cb}
		\end{subfigure}
		\begin{subfigure}[normla]{0.49\textwidth}
				\includegraphics[scale=0.26]{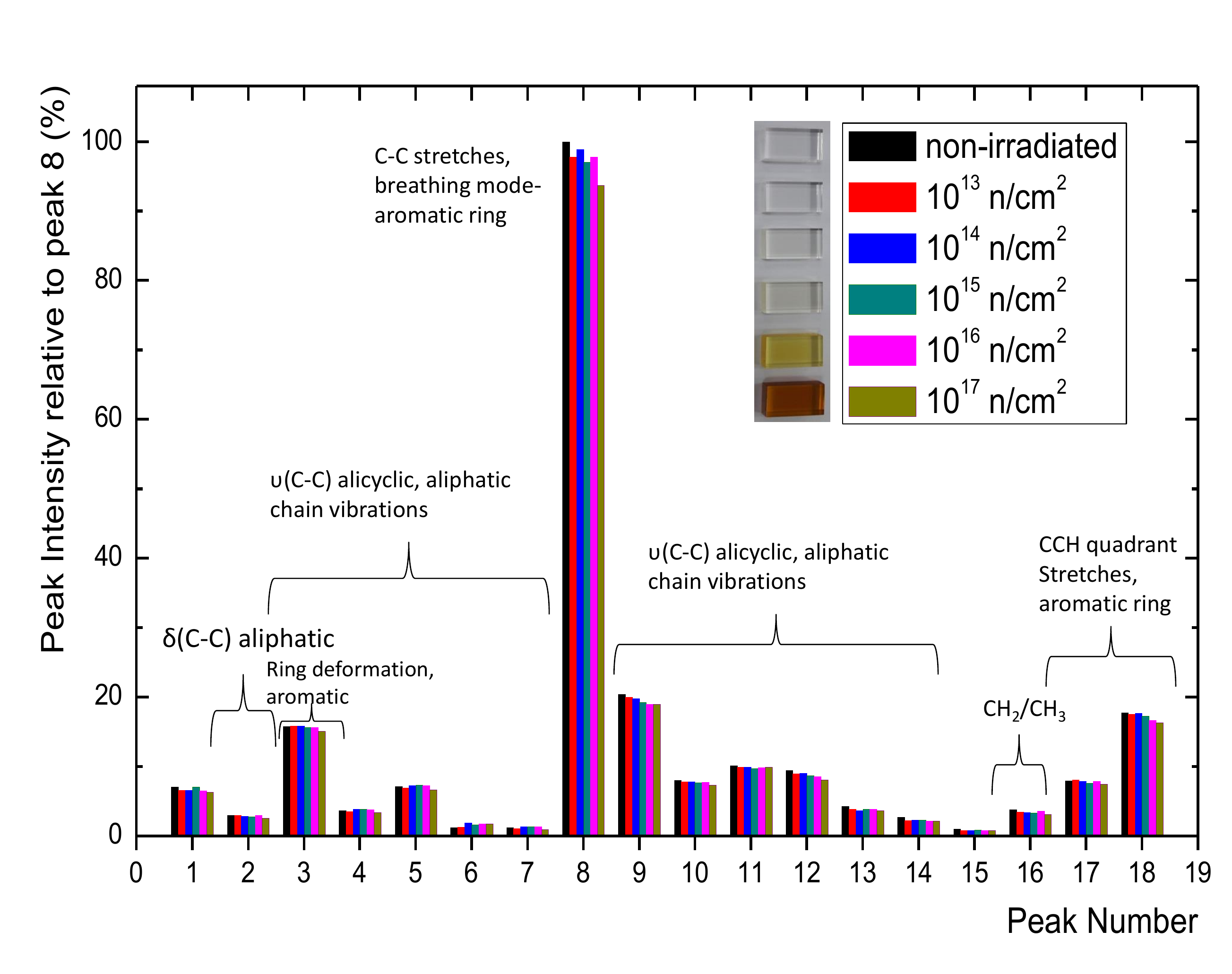}
				\caption{}
				\label{subfig:db}	  
		\end{subfigure}
		\caption{Background subtracted Raman spectra of irradiated and non-irradiated plastic scintillator samples for,~\subref{subfig:aramspec} green plastic scintillators and~\subref{subfig:bramspec} blue plastic scintillators UPS-923A. Plot of intensities of peaks relative to peak 12 (non-irradiated) for green~\subref{subfig:cb} and peak 8 (non-irradiated) for blue scintillators~\subref{subfig:db}. NB: The Raman spectra in (a) and (b) have been vertically offset for better visual presentation.}
	\end{center}
\end{figure}

The only limitation of the 785 nm wavelength is that due to detector limitations on the instrument, Raman peaks can only be measured up to 2000~cm$^{-1}$. The -C-H and =C-H vibrational modes between 2800 and 3200~cm$^{-1}$ could thus not be measured. Raman spectroscopy showed small changes in the structure, in comparison with non-irradiated and irradiated samples. Figure~\ref{subfig:aramspec} and~\subref{subfig:bramspec} reports the background subtracted Raman spectra of the green and blue emitting plastic scintillators .

Investigations were conducted in the quest of determining the radiation sensitive peaks. Intensities of the Raman peaks for both green and blue emitters were plotted relative to peak 12 and 8, respectively, in order to assess changes in the species present. Peak 12 and 8 typically represent aromatic ring structures, which influences the scintillation properties of the scintillator. The C-C bonds present in the structure give rise to a cloud of delocalized $\pi$-electrons that are prone to excitation by incident energetic particles. The results are shown in Figure~\ref{subfig:cb} and~\subref{subfig:db}. The ratio of most species found in the styrene backbone of samples to that found in the benzene ring shows a decrease after irradiation. 
   
Some of the important molecular vibrational assignments are provided in Table~\ref{table:assignments}. Peaks are assigned to their corresponding vibrational groups~\cite{spectro, menezess}. 

\begin{table}[H]
	\begin{center}
	\caption{Raman peak number and vibrational assignments for key features in green scintillators and blue scintillators UPS-923A.}
	\label{table:assignments}
	 \begin{tabular}{llr}\hline
	 	    \multicolumn{2}{l}{Peak number}	 & 
	 	    \multicolumn{1}{r}{Assignment of vibrational mode}\\ \cline{1-2}
	 	    Green & Blue \\ \hline
	 	      1     & 1 & $\tau$(CH$_3$)\\ 
	 	      2-4 & 2 & $\delta$(C-C) aliphatic\\
	 	      5, 8-11, 13-17 & 4-7, 9-14 & $\nu$(C-C) alicyclic or aliphatic chain vibration\\
	 	      12 & 8 &C-C stretches,  breathing mode- aromatic rings\\
	 	      6, 7&3& ring deformation mode, aromatic\\
	 	      18 & 15 & $\delta$(CH$_3$)\\
	 	      19 & 16 & $\delta$(CH$_2$) or $\delta$(CH$_3$) asymmetric\\
	 	      20& & --\\
	 	      21& 17& C-C aromatic stretching\\  
	 	      22 & 18 &CCH quadrant stretches- aromatic rings\\
	 	      23&&--\\
	 	       \hline 	      
   	 \end{tabular}	
	\end{center}
\end{table}

From the results shown in Figure~\ref{subfig:cb} and~\subref{subfig:db}, there is a significant decrease in the Raman intensities of aromatic breathing modes as well as alicyclic/aliphatic chain vibrations belonging to the aromatic benzene ring. These changes are more prominent in green emitting samples. It appears that the benzene ring structure undergoes a significant amount of damage. This could be a result of strong dehydrogenation due to the C-H bond breaking in the benzene ring and an emission of different C$_{x}$H$_{y}$ groups with the absorbed neutron fluence as described by~Torris, 2002~\cite{torrisi}. This causes a decline in the number of Raman active modes and hence could account for the observable effects. The Raman intensities of corresponding vibrational modes have been quantified and reported in Table~\ref{table:green int} and~\ref{table:blue int} for green and blue emitting samples, respectively.

\begin{table}[H]
	\begin{center}
		\caption{Raman intensity values of green emitting scintillators at wavenumbers 1003.6 and 1604.9 cm$^{-1}$.}
		\label{table:green int}
		\begin{tabular}{lccccr}\hline
			Functional & Peak(cm$^{-1}$) & Fluence (n/cm$^2$) & Intensity& $\%$ Int. loss\\
			group & & & (arb. units)\\ \hline
			C-C stretches& 1003.6 & non-irradiated & 9898.9\\
			breathing mode- & & 10$^{13}$ & 9779.8& 1.2\\
			aromatic ring& & 10$^{14}$ & 9779.8& 1.2\\
			& & 10$^{15}$ & 9740.1& 1.6\\
			& & 10$^{16}$ & 9621.0& 2.8\\
			& & 10$^{17}$ & 9501.9& 4.0\\ 
			CCH quadrant & 1604.9 & non-irradiated & 2197.9\\
			stretches- & & 10$^{13}$ & 2081.9&5.3\\
			aromatic ring& & 10$^{14}$ & 2032.1& 7.5\\
			& & 10$^{15}$ & 2156.5& 1.9\\
			& & 10$^{16}$ & 2082.5& 5.3\\
			& & 10$^{17}$ & 2032.7& 7.5\\ \hline	
	   \end{tabular}	
	\end{center}
\end{table}

\begin{table}[H]
	\begin{center}
		\caption{Raman intensity values of blue emitting samples at wavenumbers 1003.6 and 1604.9 cm$^{-1}$.}
		\label{table:blue int}
		\begin{tabular}{lcccr}\hline
			Functional& Peak(cm$^{-1}$) & Fluence (n/cm$^2$) & Intensity& $\%$ Int. loss\\
			group& & &(arb. units)\\ \hline
			C-C stretches & 1003.6 & non-irradiated & 11550.4\\
			breathing mode- & & 10$^{13}$ & 11294.4& 2.2\\
			aromatic ring& & 10$^{14}$ & 11422.4& 1.1\\
			& & 10$^{15}$ & 11208.9& 2.9\\
			& & 10$^{16}$ & 11294.4& 2.2\\
			& & 10$^{17}$ & 10824.8& 6.3\\
			CCH quadrant& 1604.9 & non-irradiated & 2053.9\\
			stretches- & & 10$^{13}$ & 2021.7&1.6\\
			aromatic ring& & 10$^{14}$ & 2045.8& 0.4\\
			& & 10$^{15}$ & 1997.6& 2.7\\
			& & 10$^{16}$ & 1925.1& 6.3\\
			& & 10$^{17}$ & 1884.9& 8.2\\ \hline		
		\end{tabular}	
	\end{center}
\end{table}
 
\begin{figure}[H]
	\begin{center}
		\begin{subfigure}[normla]{0.49\textwidth}
			\includegraphics[scale=0.4]{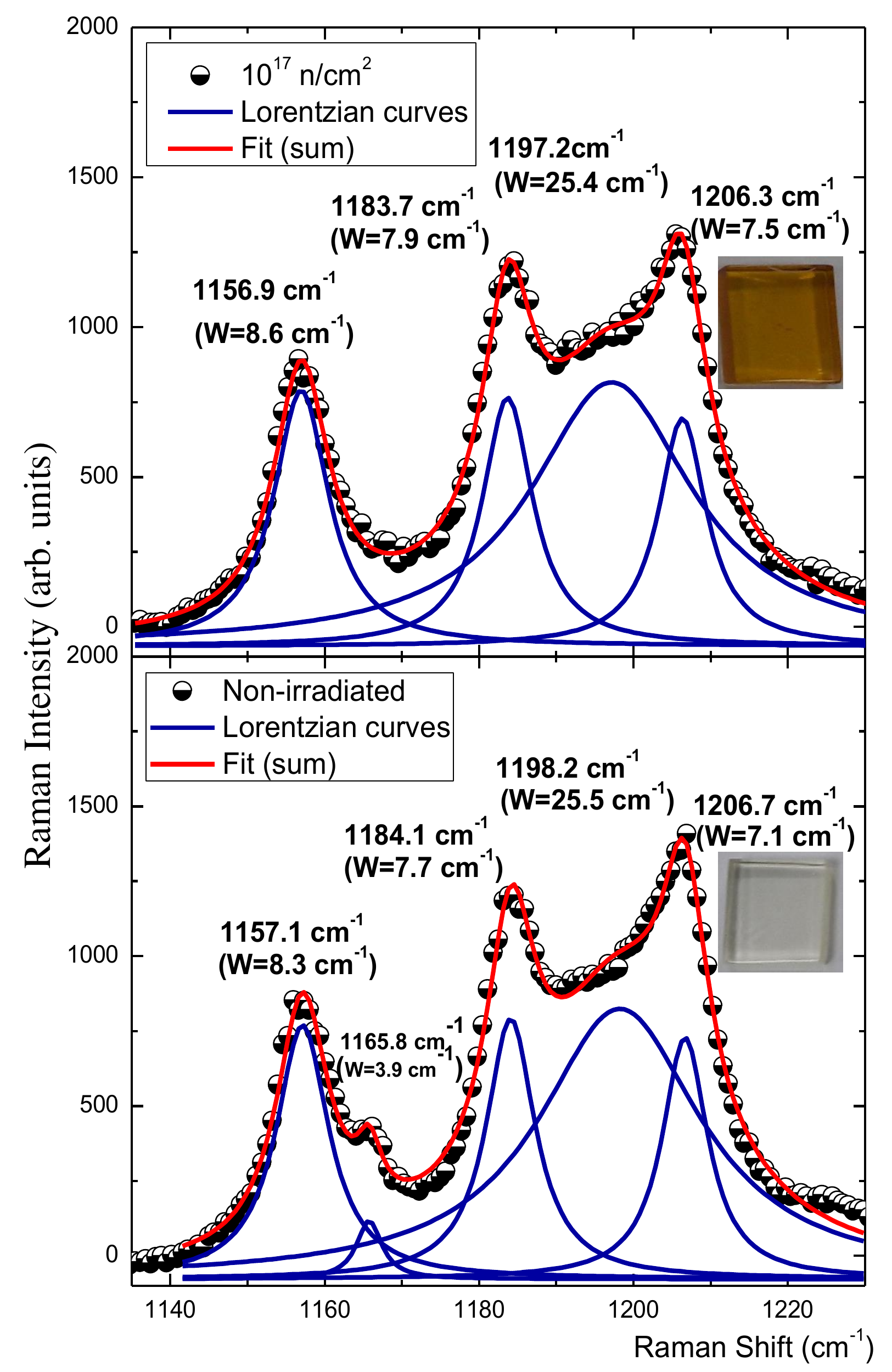}
			\caption{}
			\label{subfig:lorea}
		\end{subfigure}
		\begin{subfigure}[normla]{0.49\textwidth}
			\includegraphics[scale=0.4]{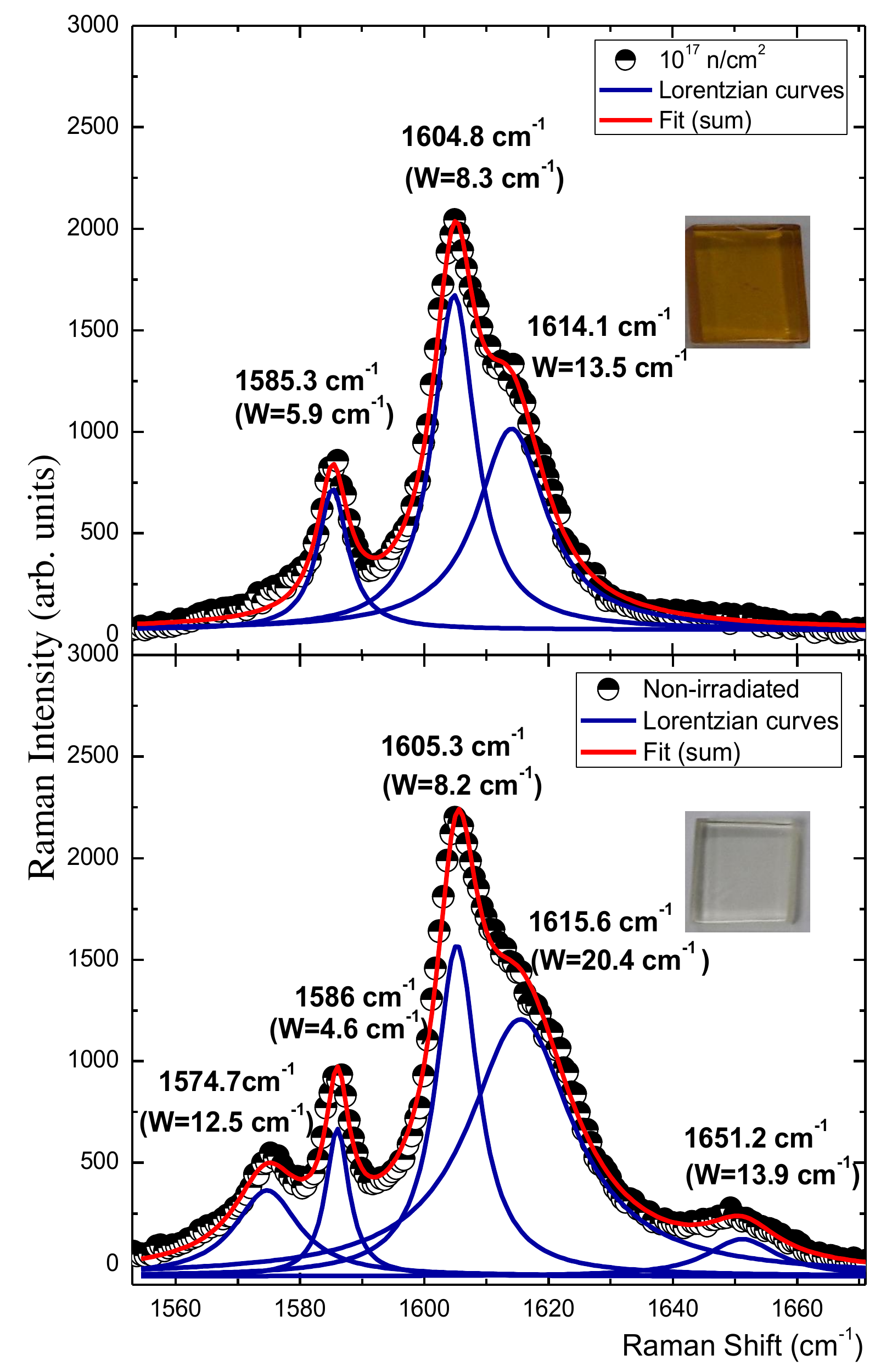}
			\caption{}
			\label{subfig:loreb}	  
		\end{subfigure}
		\caption{Lorentzian line-shape fitted function at regions 1140-1230 cm$^{-1}$~\subref{subfig:lorea} and 1560-1662 cm$^{-1}$~\subref{subfig:loreb}, showing the disappearance of Raman modes at 1165.8, 1574.7 and 1651.2 cm$^{-1}$, which are present in the non-irradiated Raman spectrum  of green scintillators.}
	\label{fig:lorentz}	
	\end{center}
\end{figure}

It has been observed that, for green scintillators, Raman peak features at frequencies 1165.8, 1574.7 and 1651.2 cm$^{-1}$ appear to be more sensitive to neutron radiation. These bands become less intense as the neutron fluence increases and finally disappear. They are most likely to be related to the fluor dopant 3HF (3-hydroxyflavone)~\cite{velmo, yu}, although the primary peaks of 3HF do not match up that well with the peaks in the green scintillators spectrum~\cite{teslova}. It could be that the fluors are a ``mix'' of 3HF derivatives. Clearly the neutron irradiation does not have a major impact on the polystyrene since all the peaks are still present even for higher fluences. The corresponding Lorentzian fits of the non-irradiated and higher neutron fluence spectra are shown in Figure~\ref{fig:lorentz} and the peak intensity values are reported in Table~\ref{table:peaks int}.

Several possible explanations for these effects can be made. The coupling interaction between the individual vibrations can result in the formation of a combined band~\cite{spectro}. A slight broadening of the closest peaks due to formation of the new chemical bonds might have also contributed to the observed changes. Sonkawade {\it{et al}}.~\cite{sonkawade} observed new bands formation ascribed to cross-linking of the polymer chains after neutron irradiation of polyaniline. Furthermore, chain scission and dehydrogenation of the polymer as described by Evans {\it{et al}}.~\cite{evans}, could be the cause of structural alterations. These support the findings reported in this paper. 
         
The luminescence properties of the fluors depend on their $\pi$-electron systems staying intact, and that if those are damaged then the fluorescence properties are adversely affected. At the moment there is not enough literature evidence to make a firm link between the actual peak and damage to the $\pi$-electron systems.

\begin{table}[H]
	\begin{center}
		\caption{Raman intensity values of green scintillators at wavenumbers 1165.8, 1574.7 and 1651.2 cm$^{-1}$.}
		\label{table:peaks int}
		\begin{tabular}{lcccr}\hline
		   Peak(cm$^{-1}$) & Fluence (n/cm$^2$) & Intensity& $\%$ Int. loss\\
		                   &               & (arb.units)\\ \hline
			     1165.8    & non-irradiated& 166.6 \\			
			               & 10$^{13}$     & 164.4 & 1.3\\
			               & 10$^{14}$     & 107.8 & 35.3\\
			               & 10$^{15}$     & 126.3 & 24.2 \\
			               & 10$^{16}$     & 130.7 & 21.5 \\
			               & 10$^{17}$     & 0     & 100\\ 	
		         1574.7    & non-irradiated& 319.1\\
	     	               & 10$^{13}$     & 263.3 & 17.5\\
			               & 10$^{14}$     & 131.7 & 58.7\\
			               & 10$^{15}$     & 175.8 & 44.9\\
			               & 10$^{16}$     & 111.8 & 64.9\\
			               & 10$^{17}$     & 0     & 100\\ 	
			     1651.2    & non-irradiated& 141.8\\
			               & 10$^{13}$     & 99.9  & 29.5\\
			               & 10$^{14}$     & 72.5  &48.9\\
			               & 10$^{15}$     & 70.9  & 50.0\\
			               & 10$^{16}$     & 19.3  & 86.4\\
			               & 10$^{17}$     & 0     & 100\\ \hline				
		\end{tabular}	
	\end{center}
\end{table}

\subsection{Fluorescence Spectroscopy Results and Analysis}
\begin{figure}[H]
	\begin{center}
		\begin{subfigure}[normla]{0.49\textwidth}
			\includegraphics[scale=0.26]{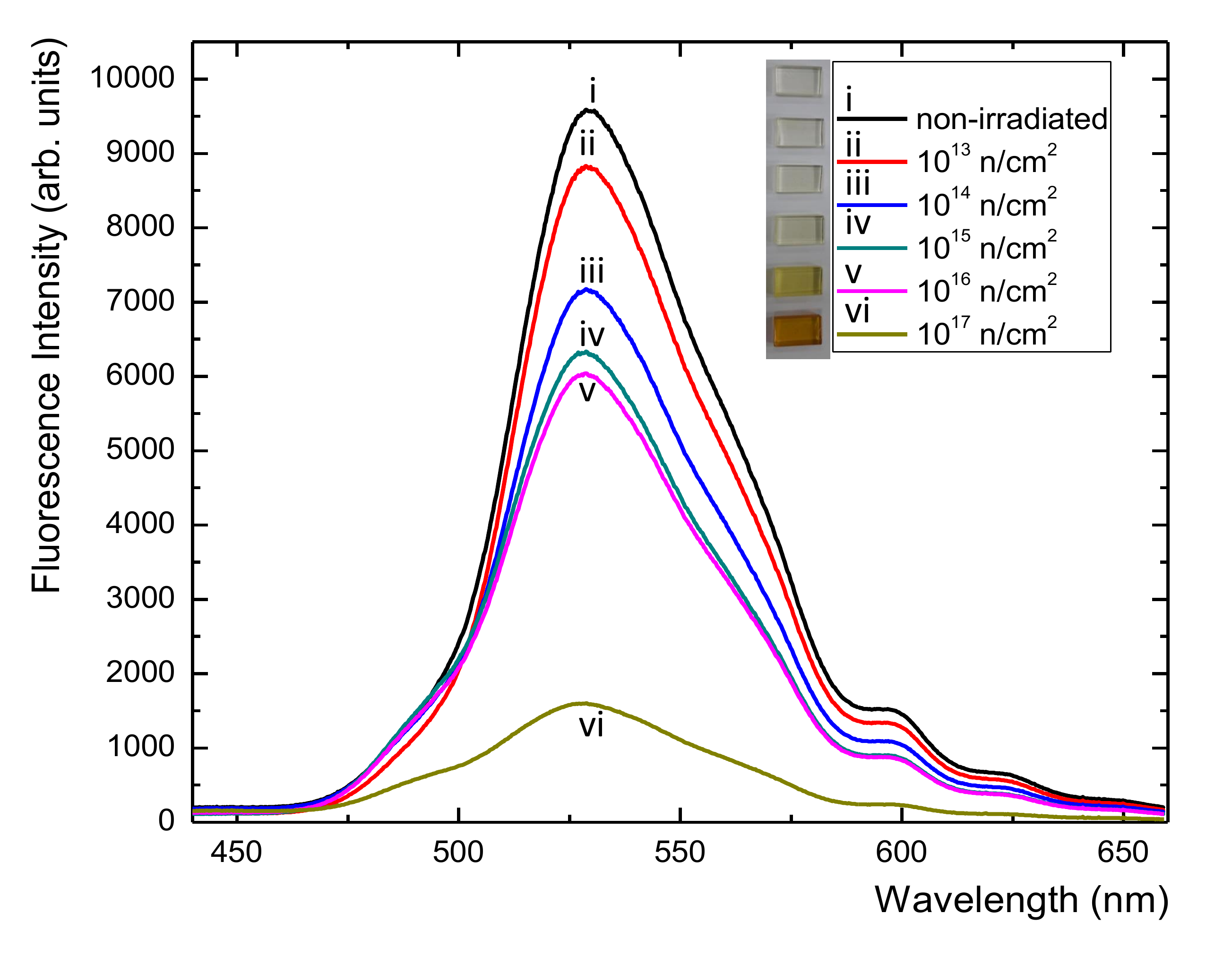}
			\caption{}
			\label{subfig:af}
		\end{subfigure}
		\begin{subfigure}[normla]{0.49\textwidth}
			\includegraphics[scale=0.26]{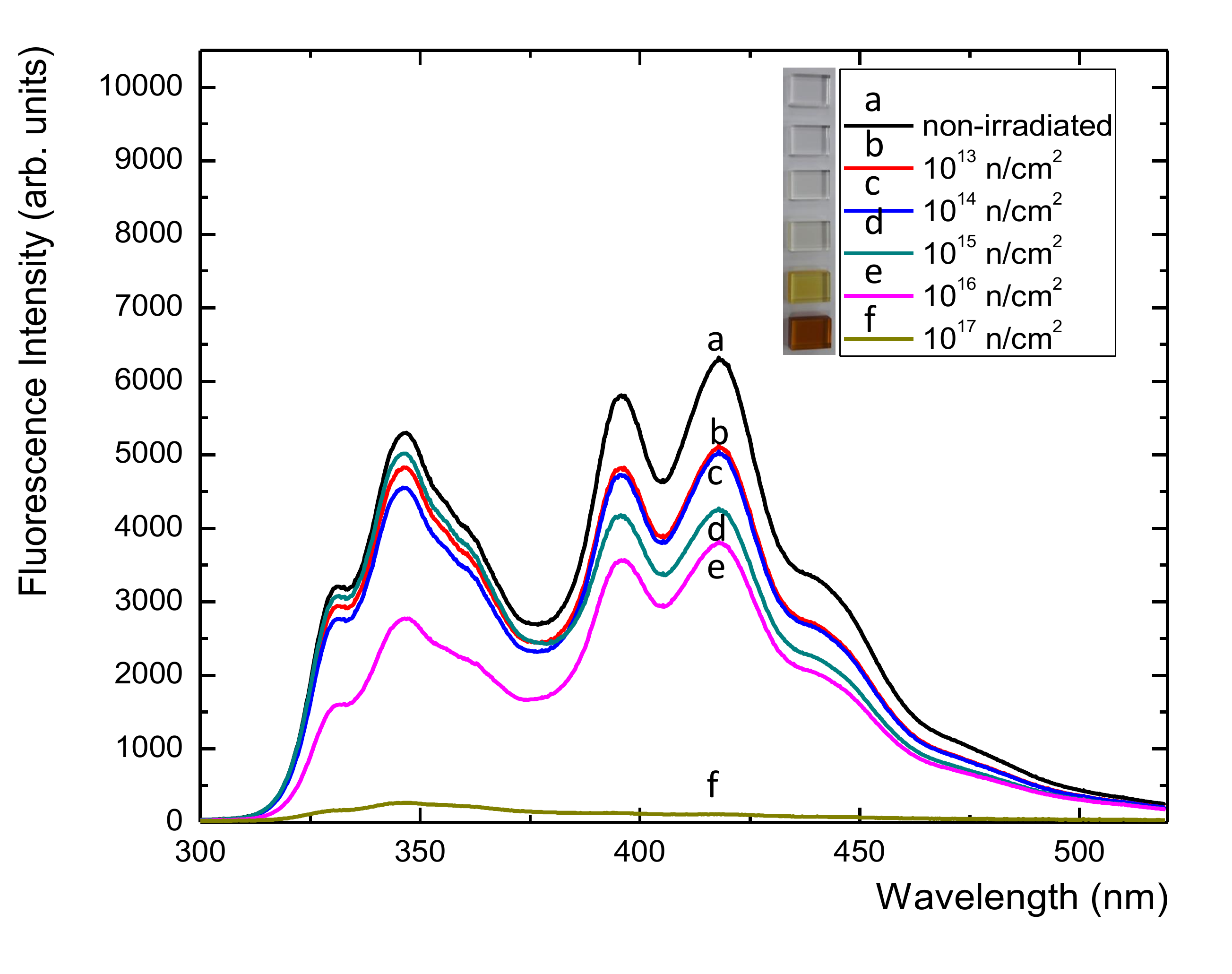}
			\caption{}
			\label{subfig:bf}	  
		\end{subfigure}
		\caption{Fluorescence spectra for green scintillators~\subref{subfig:af} and blue UPS-923A scintillators~\subref{subfig:bf}.}
     	\label{fig:fluorescence}
\end{center}

\end{figure}

Fluorescence spectra of plastic scintillator samples are shown in Figure~\ref{fig:fluorescence}. A two-peak feature in Figure~\ref{subfig:bf} is observed for blue scintillators UPS-923A. Wavelength regions of 310-375 nm and 375-520 nm correlate with fluorescence of the polystyrene base and fluor dopants, respectively. The two-peak feature is observed since the fluorescence is predominantly from the benzene ring structures. However, the results for green emitting scintillators in Figure~\ref{subfig:af} reveal only a single fluorescence peak around 529 nm. For all samples, the decrease in fluorescence intensity is prominent as neutron fluence increases. There could be degradation of the aromatic benzene ring structures in the polymer base matrix and damage to the f\"{o}rster energy transfer mechanisms due to C-H bond breaking within the benzene ring. These damages are reported in literature by Torrisi, 2002~\cite{torrisi}, to be the cause of luminescence yield reduction. The results of Raman spectroscopy reported in Table~\ref{table:green int} and~\ref{table:blue int} affirm these changes, and are consistent with the fluorescence results. Damage to the benzene ring directly affects the scintillation process in the material.

\subsection{Light Transmission Results and Analysis}
\begin{figure}[H]
	\begin{center}
		\begin{subfigure}[normla]{0.49\textwidth}
			\includegraphics[scale=0.265]{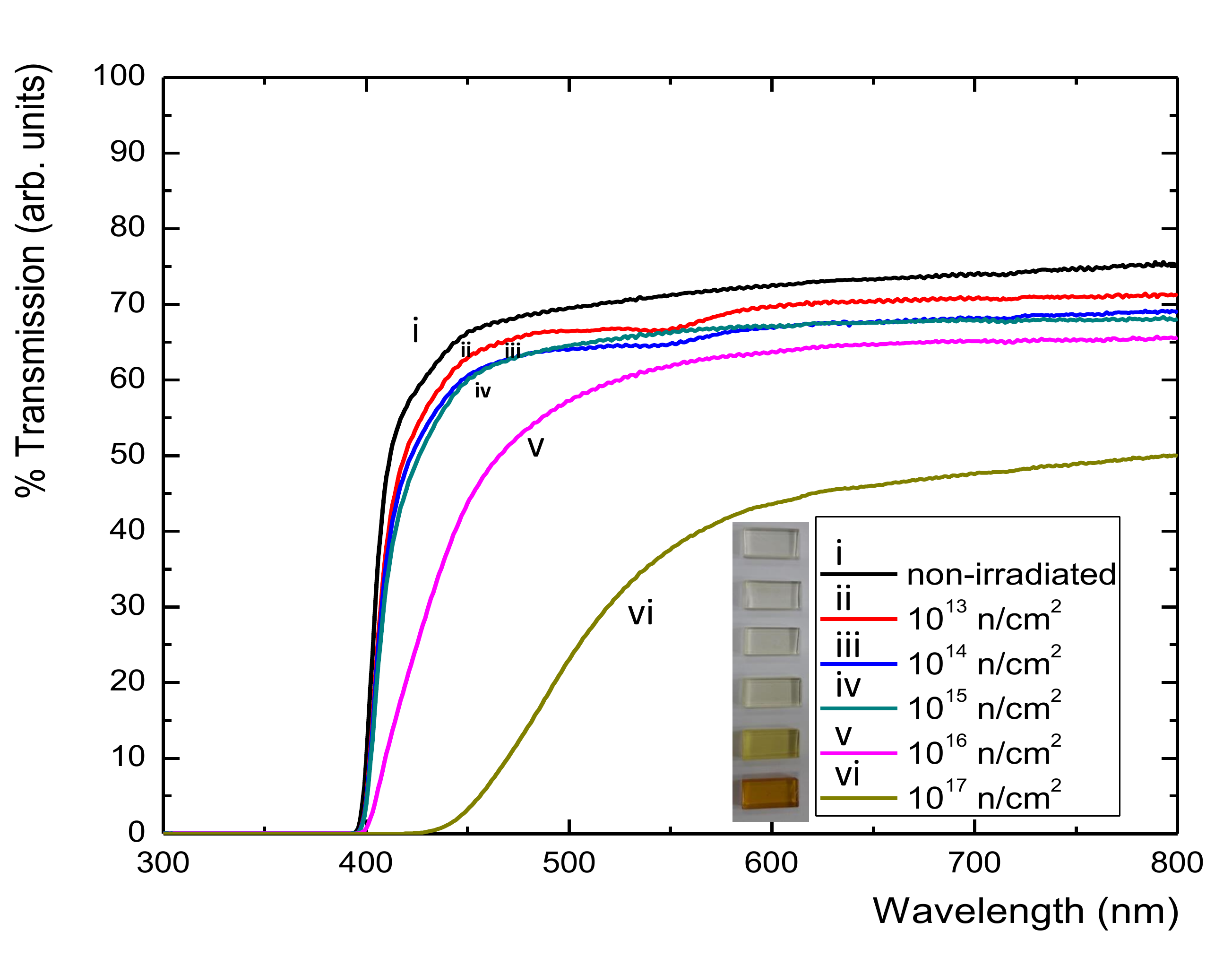}
			\caption{}
			\label{subfig:at}
		\end{subfigure}
		\begin{subfigure}[normla]{0.49\textwidth}
			\includegraphics[scale=0.265]{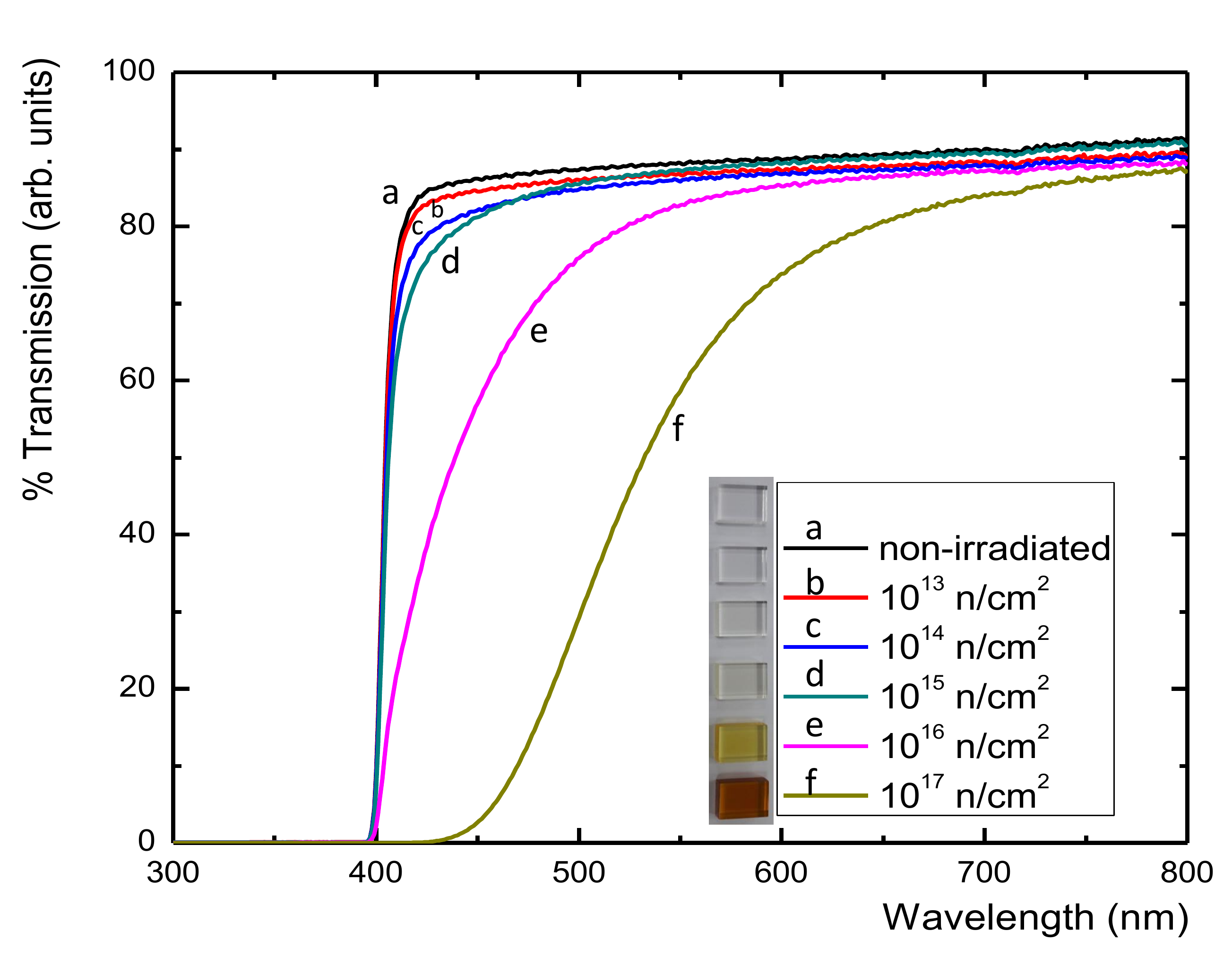}
			\caption{}
			\label{subfig:bt}	  
		\end{subfigure}
		\caption{Light transmission spectra for green scintillators~\subref{subfig:at} and blue scintillators UPS-923A~\subref{subfig:bt}.}
		\label{fig:trans}
	\end{center}
\end{figure}

Transmission spectroscopy was conducted using the Varian Cary 500 spectrophotometer with the light transmission measured relative to transmission in air over a range of 300-800 nm. The results are shown in Figure~\ref{fig:trans}. All the non-irradiated blue scintillators UPS-923A and green scintillators have an absorption edge starting at ~410 nm, which completely falls off at around 400 nm. The formation of an absorptive tint as neutron fluence increases is observed where the absorption edge drops and shifts to longer wavelengths. This effect is ascribed to the production of free radicals induced by radiation damage. These free radicals form absorption centers within the samples resulting in the light absorption competition and loss of transparency in scintillators. 

From the light transmission plots, it is evident that blue scintillators UPS-923A possess the highest light transmission properties and appear to be most radiation tolerant as compared to green light emitters. The blue scintillator overall remains more transparent to all wavelengths, but becomes opaque to blue light after damage.  Whereas, the green scintillators lose transmission in blue light region, but remain transparent in the 550+ nm region.

\subsection{Light Yield Results and Analysis}
\begin{figure}[H]
	\begin{center}
		\begin{subfigure}[normla]{0.49\textwidth}
			\includegraphics[scale=0.44]{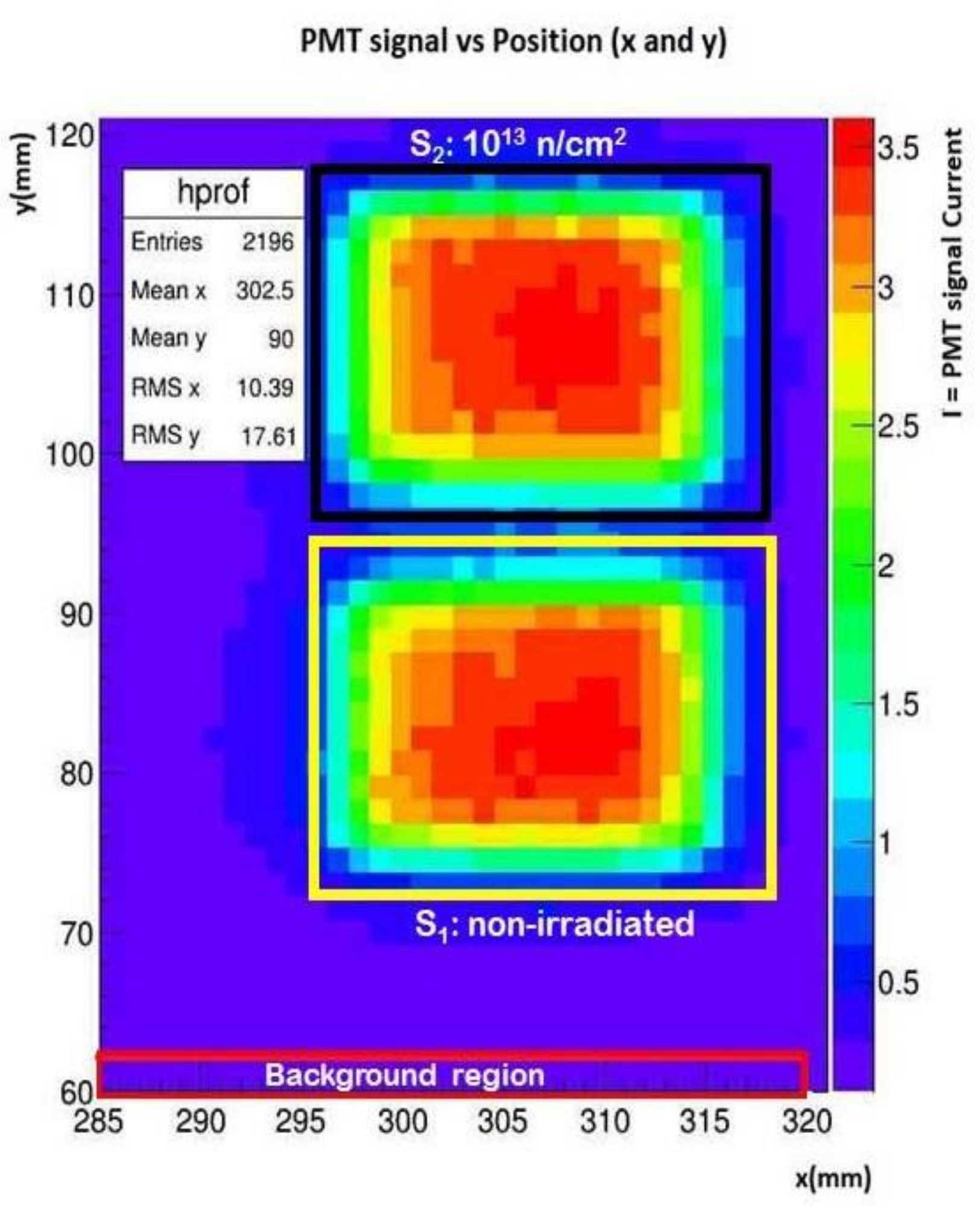}
			\caption{}
			\label{subfig:back}
		\end{subfigure}
		\begin{subfigure}[normla]{0.49\textwidth}
			\includegraphics[scale=0.44]{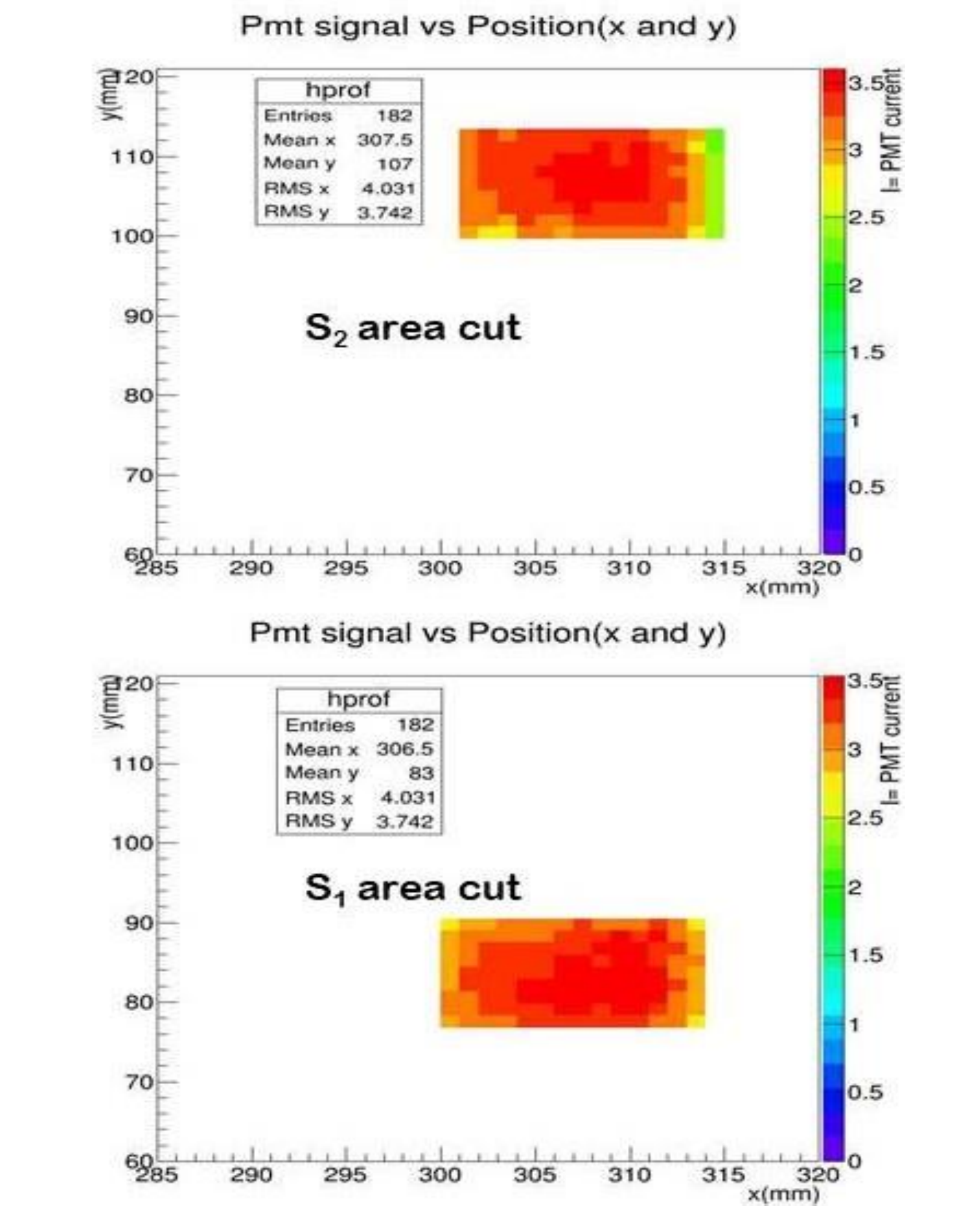}
			\caption{}
			\label{subfig:area}	  
		\end{subfigure}
		\caption{2D mapping of the PMT signal with \ce{^{90}Sr} source position, indicating signal regions corresponding to regions on the experimental set-up~\subref{subfig:back} and plots of area cut with entries of high signal values~\subref{subfig:area} [color legend only available on electronic version].}
	\end{center}
\end{figure}

The assessment of light yielded by irradiated plastic scintillator samples was performed by testing their response to a \ce{^{90}Sr} $\beta$-electron source. The signal generated by the PhotoMultiplier Tube was measured as a function of radiation source position. Figure~\ref{subfig:back} illustrates a 2D mapping of the signal measured in the X and Y direction over a pair of samples. The position of samples is approximated by the black and yellow boxes (color legend only available on electronic version), representing irradiated and non-irradiated samples, respectively.            

Before analysis, the background was subtracted from the rest of the mappings. The background region marked with a red box on Figure~\ref{subfig:back} was selected on one of the 2D mapping surface with the signal not contaminated by the signal from the samples. A pair of irradiated (10$^{13}$ n/cm$^2$) blue test sample and the non-irradiated blue reference sample were used. The mean signal value of 0.1309 mean/entry over the selected area was used as a background value. 

\begin{figure}[H]
	\begin{center}
		\begin{subfigure}[normla]{1\textwidth}
			\includegraphics[height=8cm, width=14cm]{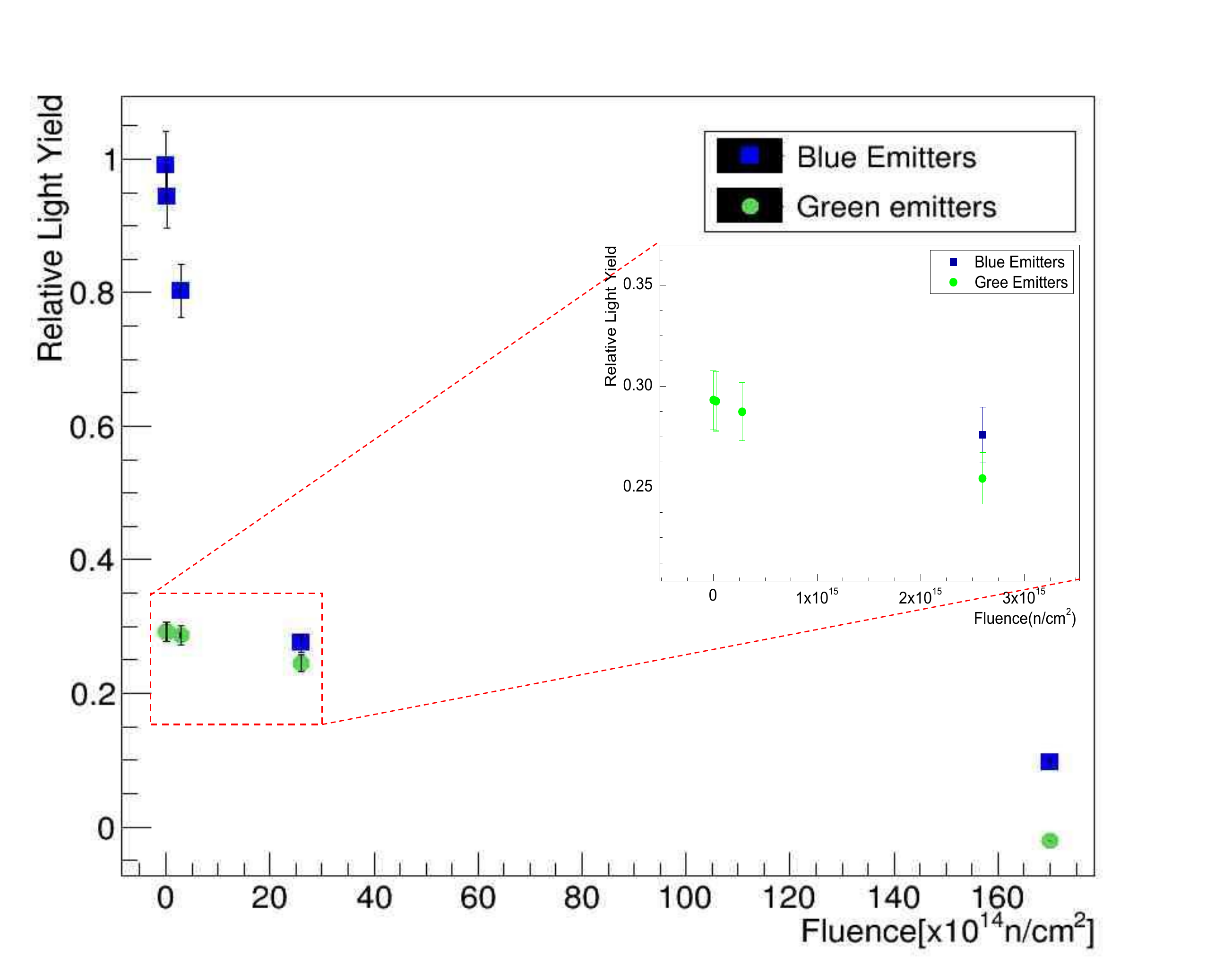}
		\end{subfigure}
		\caption{Light yield against neutron fluence for blue UPS-293A scintillator and green scintillator.}
		\label{fig:rly}
	\end{center}
\end{figure}

The area cuts plots shown in Figure~\ref{subfig:area} were selected from the 2D mappings with entries of high signal values. In order to analyse the light yield loss as a function of neutron fluence, the ratio of corrected signals from the reference and test samples were calculated and plotted against the neutron fluence as shown in Figure~\ref{fig:rly}. The light yield loss in all irradiated samples is observed as neutron fluence increases. It is prominent at a fluence of 10$^{17}$ n/cm$^2$. The light yield results are consistent with the fluorescence and light transmission results.     

The radiation damage experienced by plastic scintillators may be influenced by a number of factors such as the total dose absorbed and dose-rate. The influence of dose rate in predicting the lifetime of plastic scintillators has not yet been studied with neutron radiation.  On the other hand, radiation damage studies have been conducted using proton beams provided by the 6 MV EN Tandem accelerator at iThemba LABS, Gauteng. Several dose rates were used and it was observed that plastic scintillators exposed to high dose rates degraded more in comparison to a low dose rates.

\section{Conclusion}
The effects of neutron radiation on the structural and optical properties of blue scintillators UPS-923A and green scintillators were studied. According to the results obtained, irradiation to high neutron fluences with energy of $E>1$ MeV, marginally affected the structural properties and strongly affected the optical properties. The effects of neutron irradiation on the Raman spectroscopy, fluorescence, light transmission and light yield at different fluences have been demonstrated. A discolouration in the samples emerged. It is prominent at a fluence of 10$^{16}$ n/cm$^2$ and continues with an increase in fluence. This is attributed to the formation of free radicals due to dehydrogenation induced by the neutron bombardment. These free radicals combine to produce cross-links which result in the formation of a three dimensional network and cause discolouration. Raman spectroscopy revealed that blue scintillators UPS-923A maintain their structural characteristics after irradiation with slight intensity loss in some species. However, for green scintillators, it is pointed that Raman peak features at frequencies 1165.8, 1574.7 and 1651.2 cm$^{-1}$ appear to be more radiation sensitive and die out with an increase in neutron fluence.

Radiation damage decreases the transmittance of light, the luminescence intensity, and the light yield at relatively high fluences. The optical properties are altered by the presence of radiation induced radicals. These excited species form when bonds break within the polymer. Free radicals initiate chemical reactions that alter the structure of the polymer backbone of the plastic. More effects of neutron damage is observed as irradiation progresses to high fluences. Furthermore, the degradation of the polymer base matrix which results in the damage of $\pi$-electron structure in the benzene ring largely contribute to the observed modifications.

\section{Acknowledgements} 
The authors are grateful to staff of the Institute for Scintillation Materials (ISMA) in Kharkov, Ukraine for providing plastic scintillators for this study, the technical team at the IBR-2 reactor of the Frank Laboratory of Neutron Physics in Dubna, Russia for providing neutron irradiation, the University of the Witwatersrand for making the equipment available for analysis, as well as Dr. Oleg Solovianov for his help with the light yield measurements at CERN. This project was funded by BIUST, SA-CERN consortium, NRF-SA and SA-JINR.


\section{References}
\begin{thebibliography}{00}

\bibitem[Knoll(1999)]{Knoll:1300754}
G. F.~Knoll, 3$^{rd}$, Radiation Detection and Measurement, John Wiley $\&$ Sons Inc, Michigan, 1999 (Chapter 8, pp. 220-222)

\bibitem[Chen(2011)]{chen}
M.~Chen, Queen's University, PHYS 352: Measurement, Instrumentation and Experiment Design [Online]. Available: http://www.physics.queensu.ca/~phys352/lect19.pdf, 2011.

\bibitem{Angela}
A. Vasilescu, Overview on the radiation environment in ATLAS and CMS SCT and
the irradiation facilities used for damage tests. ROSE/TN/97-3, 1997.

\bibitem{sonkawade}
R. G.~Sonkawade {\it et al.}, 
Effects of gamma ray and neutron radiation on polyaniline conducting polymer. Indian Journal of Pure $\&$ Applied Physics. 48 (2010) 453-456

\bibitem{SAIP}
H.~Jivan {\it et al.},
Radiation hardness of plastic scintillators for the Tile Calorimeter of
the ATLAS detector, Proceedings of SAIP2014, University of Johannesburg, 978-0-
620-65391-6, 2014, pp. 199-205.

\bibitem{1742-6596-645-1-012019}
H.~Jivan {\it et al.},
Radiation hardness of plastic scintillators for the Tile Calorimeter of the ATLAS detector, J. Phys.: Conf. Seri. 645 (1) (2015), https://doi.org/10.1088/
1742-6596/645/1/012019.

\bibitem{NIMB}
H.~Jivan {\it et al.},
Radiation damage effects on the optical properties of plastic scintillators. Nucl. Instrum. Methods Phys. Res. Section B Beam Interactions Mater. At. 409 (2017) 224-228.

\bibitem{Bisanti}
P. Bisanti, F. Borsa, V. Tognetti (Eds.), Magnetic properties of matter, World
Scientific, Turin, 1986, pp. 385-408.

\bibitem{sci}
Institute for Scintillation Materials (ISMA), Available: <http://www.isma.kharkov.ua/eng/>

\bibitem{velmo}
E.S.~Velmozhnaya {\it et al.},
The new radiation-hard plastic scintillators with diffusion
enhancers and 3-hydroxyflavone derivatives, Funct. Mater. 23 (4) (2006) 650-656.

\bibitem{yu}
Yu.A.~Gurkalenko {\it et al.},
The plastic scintillator activated with fluorinated 3-hydroxyflavone, Funct. Mater. 24 (2) (2017) 244-249.

\bibitem{bulav}
M.~Bulavin {\it et al.},
Irradiation facility at the IBR-2 reactor for investigation of material
radiation hardness, Nucl. Instrum. Methods B 343 (2015) 26-29.

\bibitem{shabalin}
E. P.~Shabalin {\it et al.}, 
Spectrum and density of neutron flux in the irradiation beam
line no. 3 of the IBR-2 reactor, Phys. Particles Nucl. Lett. 12 (2) (2015) 336-343.

\bibitem{spectro}
R. M.~Silverstein {\it et al.},
Spectrometric identification of organic compounds, 7th ed,
John Wiley $\&$ Sons Inc., New York, 2005 (Chapter 2, pp 82).

\bibitem{menezess}
D. B.~Menezess {\it et al.}, Glass transition of polystyrene (PS) studied by Raman spectroscopic investigation of its phenyl functional groups, Mater. Res. Express 4 (1) (2017) 015-303.

\bibitem{torrisi}
L.~Torrisi, 
Radiation damage in polyvinyltoluene (PVT), Radiation Phys. Chem. 63 (2002) 89-92.

\bibitem{teslova}
T.~Teslova {\it et al.},
Raman and surface-enhanced Raman spectra of flavone and several hydroxy derivatives, J. Raman Spectroscopy 38 (7) (2007) 802-818.

\bibitem{evans}
D.~Evans {\it et al.},
Irradiation of plastics: damage and gas evolution, MRS Bull. 22 (1997) 36-40.
 

\end{thebibliography}
\end{document}